\documentclass[a4paper,10pt]{article}
\usepackage[T2A]{fontenc}
\usepackage{amssymb,amsfonts,amsmath,mathtext,cite,enumerate,float,mathrsfs}
\usepackage[utf8]{inputenc}
\usepackage[english]{babel}
\usepackage{graphicx,color}
\usepackage{wrapfig}
\graphicspath{{./}}

\usepackage{epsfig}
\usepackage{epsf}
\usepackage{subfig} 
\usepackage{feynmp} 
\usepackage{pgf,pgfarrows,pgfnodes,pgfautomata,pgfheaps,multirow}
\usepackage{array}
\usepackage{bm}
\usepackage{latexsym}
\usepackage{pifont}

\newcommand{\be}{\begin{equation}}
\newcommand{\ee}{\end{equation}}
\newcommand{\bea}{\begin{eqnarray}}
\newcommand{\eea}{\end{eqnarray}}
\newcommand{\dd}{\mbox{d}} 
\newcommand\nn{\nonumber}

\title{The decays $\rho^{-}\to\eta\pi^{-}$ and $\tau^{-}\to\eta(\eta')\pi^{-}\nu$ in the NJL model}

\author{M. K. Volkov\footnote{E-mail address: volkov@theor.jinr.ru}, 
D. G. Kostunin\footnote{E-mail address: kostunin@theor.jinr.ru}\\
\it Bogoliubov Laboratory of Theoretical Physics, JINR\\ 
\it Dubna, 141980, Russia}

\begin{document}

\maketitle

\begin{abstract}
The widths of the decays $\rho^{-}\to\eta\pi^{-}$ and $\tau^{-}\to\eta(\eta')\pi^{-}\nu$ are calculated in the framework of the NJL model.
It is shown that these decays are defined by the $u$ and $d$ quark mass difference. 
It leads to the suppression of these decays in comparison with the main decay modes.
In the process $\rho^{-}\to\eta\pi^{-}$ the intermediate scalar $a_0^{-}$ state is taken into account.
For the $\tau$ decays the intermediate states with $a_0^{-}$, $\rho^{-}(770)$ and $\rho^{-}(1450)$ mesons are used.
Our estimates are compared with the results obtained in other works.
\\

\\


12.39.Fe 	Chiral Lagrangians 
\end{abstract}



\section{Introduction}
At present, the decays $\rho^{-}\to\eta\pi^{-}$ and $\tau^{-}\to\eta(\eta')\pi^{-}\nu$ are not well studied in experiments~\cite{PDG,Ferbel:1966zz,delAmoSanchez:2010pc}.
However, recently, a number of works devoted to the investigation of these processes in the framework of the different phenomenological models were published~\cite{Tisserant:1982fc,Bramon:1987zb,Neufeld:1994eg,Nussinov:2008gx,Nussinov:2009sn,Paver:2010mz,Paver:2011md}.
On the other hand, in~\cite{VolkovIvanovOsipov3pi, VolkovIvanovOsipovpigamma,Volkov:2012uh,Volkov:2012gv} it was shown that different modes of the $\tau$ decay can be satisfactorily described in the NJL model~\cite{VolkovEbert, VolkovAn, pepan86, EbertReinhardt, pepan93, VolkovEbertReinhardt, UFN}.
In the present paper, the NJL model is used for the description of the decays $\rho^{-}\to\eta\pi^{-}$ and $\tau^{-}\to\eta(\eta')\pi^{-}\nu$.
The probabilities of the transitions $\pi^{0} \to \eta(\eta')$ and $\rho^{-}(W^{-}) \to a_{0}^{-}$ are calculated.
These transitions are defined by the mass difference between $u$ and $d$ quarks and can be calculated in the framework of the NJL model without attraction of any arbitrary parameters.
Our results will be compared with the estimates obtained in~\cite{Nussinov:2008gx,Nussinov:2009sn,Paver:2010mz,Paver:2011md}.
It is shown that the amplitudes with intermediate vector mesons dominate in the $\tau^{-}\to\eta\pi^{-}\nu$ decay.

\section{The decay $\rho^{-}\to\eta\pi^{-}$}
\begin{figure}[h]
 \begin{center}

 \begin{tabular}{cc}
\begin{fmffile}{etapi}
      \begin{fmfgraph*}(150,30)

	      \fmfpen{thin}\fmfleftn{l}{2}\fmfrightn{r}{2}
	      \fmfdot{f}
	      \fmfright{a}
	      \fmfleft{f}
	      \fmf{dbl_plain_arrow,lab.side=left,label=$\pi^{0}$}{f,v1}

	      \fmf{fermion,left,tension=.8}{v1,v2}
	      \fmf{fermion,left,tension=.8}{v2,v1}

	      \fmf{dbl_plain_arrow,lab.side=left,label=$\eta(\eta')$}{v2,a}
\end{fmfgraph*}
\end{fmffile}

& \begin{fmffile}{vtoa0}
      \begin{fmfgraph*}(150,30)

	      \fmfpen{thin}\fmfleftn{l}{2}\fmfrightn{r}{2}
	      \fmfright{a}
	      \fmfleft{f}
	      \fmf{dbl_plain_arrow,lab.side=left,label=$\rho^{-}(W^{-})$}{f,v1}

	      \fmf{fermion,left,tension=.8}{v1,v2}
	      \fmf{fermion,left,tension=.8}{v2,v1}

	      \fmf{dbl_plain_arrow,lab.side=left,label=$a^{-}_0$}{v2,a}

      \end{fmfgraph*}
\end{fmffile}
\end{tabular}
\caption{$\pi^{0} \to \eta(\eta')$ (left) and $\rho^{-}(W^{-}) \to a_0^{-}$ (right) transitions}
\label{fig1}
\end{center}
\end{figure}
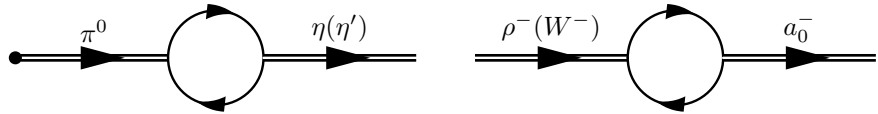

For calculation of this decay we should first calculate two non-diagonal transitions $\pi^{0} \to \eta$ and $\rho^{-} \to a_0^{-}$ within the NJL model. These transitions go through quark loops containing $u$ and $d$ quarks (see Fig.~\ref{fig1}).

The amplitude of the transition $\pi^{0} \to \eta(\eta')$ has the form
\begin{equation}
\epsilon_{\pi\eta(\eta')} = 2g_{\pi}^2 ((2I_1(m_d) + m_{\eta(\eta')}^2I_2(m_d)) - (2I_1(m_u) + m_{\eta(\eta')}^2I_2(m_u)))\frac{\epsilon_{\eta(\eta')}}{m_\pi^2 - m_{\eta(\eta')}^2}\,,
\label{pietatrans}
\end{equation}
where $m_{\pi}$, $m_{\eta}$, $m_{\eta'}$ are masses of $\pi$, $\eta$ and $\eta'$ mesons, respectively, given in PDG~\cite{PDG}; $m_u$ and $m_d$ are constituent quark masses, $m_u = 280$ MeV. Using the last experimental data for the decay $\omega \to \pi\pi$~\cite{PDG} we obtain $m_d - m_u \approx 3.7$ MeV. This decay was described in detail in~\cite{pepan86}. The $\eta$ -- $\eta'$ mixing $\epsilon_{\eta} = \sin{\bar{\theta}}$ for the $\eta$ meson and $\epsilon_{\eta'} = \cos{\bar{\theta}}$ for the $\eta'$ meson. The mixing angle $\bar{\theta} \approx -54^{\circ}$ was defined in~\cite{Volkov:1998ax}.
The constant $g_{\pi}$ and integrals $I_1(m)$, $I_2(m)$ are defined in~\cite{pepan86}
\begin{eqnarray}
g_{\pi} &=& \frac{m_u}{F_{\pi}}\,, \\
I_1(m) &=& -i\frac{N_c}{(2\pi)^4}\int^{\Lambda_4} \frac{\dd^4k}{(m^2 - k^2)} = \frac{N_c}{(4\pi)^2}\left[\Lambda_4^2 - m^2\log\left(\frac{\Lambda_4^2}{m^2}+1\right)\right]\,, \\
I_2(m) &=& -i\frac{N_c}{(2\pi)^4}\int^{\Lambda_4} \frac{\dd^4k}{(m^2 - k^2)^2} = \frac{N_c}{(4\pi)^2}\left[\log\left(\frac{\Lambda_4^2}{m^2} + 1\right) - \left(1 + \frac{m^2}{\Lambda_4^2}\right)^{-1}\right]\,,
\end{eqnarray}
where $N_c = 3$ is a number of quark colors and $\Lambda_4 \approx 1250$ MeV is a 4-dimentional cut-off parameter in the standard NJL model~\cite{pepan86}.

The obtained estimates coincide with those used in~\cite{Paver:2010mz,Paver:2011md}. One can see the comparison in Table~\ref{tbl1}.
\begin{table}[ht]
\begin{center}
\caption{Comparison of $\epsilon_{\pi\eta(\eta')}$}
{\begin{tabular}{@{}|l|l|l|l|@{}} 
 $|\epsilon_{\pi\eta}^{PR}|$~\cite{Paver:2010mz} & $|\epsilon_{\pi\eta}^{NJL}|$ & $|\epsilon_{\pi\eta'}^{PR}|$~\cite{Paver:2011md} &  $|\epsilon_{\pi\eta'}^{NJL}|$ \\
 $1.34 \cdot 10^{-2}$ & $1.55 \cdot 10^{-2}$ & ($3 \pm 1) \cdot 10^{-3}$ & $6.79 \cdot 10^3$ \\ 
\end{tabular}}
\label{tbl1}
\end{center}
\end{table}

The transition $\rho^{-} \to a^{-}_0$ takes the form
\begin{equation}
\frac{\sqrt{6}}{2}(m_d - m_u)p^{\mu}\rho_{\mu}^{-}a_0^{-}\,,
\end{equation}
where $p$ is momentum of the $\rho$ meson.

The $\rho$ meson decay width is defined by two diagrams in Figs.~\ref{fig2} and~\ref{fig3}.
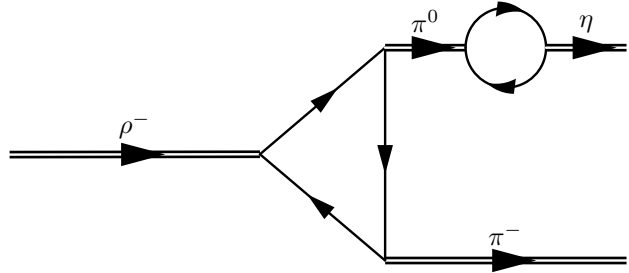
\begin{figure}[h]
\begin{center}
\begin{fmffile}{rho1}
      \begin{fmfgraph*}(250,100)
	      \fmfpen{thin}\fmfleftn{l}{2}\fmfrightn{r}{2}
 
	      \fmfright{b,a1}
	      \fmfleft{f}
 	      \fmf{fermion}{p1,v2,p2}
	      \fmf{dbl_plain_arrow,lab.side=left,label=$\rho^{-}$}{f,v2}
 	      \fmf{fermion}{p2,p1}	      

 	      \fmf{dbl_plain_arrow,lab.side=left,label=$\pi^{-}$}{p1,b}

	      \fmf{dbl_plain_arrow,lab.side=left,label=$\pi^{0}$}{p2,v5}

	      \fmf{fermion,left,tension=.5}{v5,v6}
	      \fmf{fermion,left,tension=.5}{v6,v5}

	      \fmf{dbl_plain_arrow,lab.side=left,label=$\eta$}{v6,a1}

	      \fmfforce{140,90}{p2}
	      \fmfforce{140,10}{p1}

	      \fmfforce{230,90}{a1}
	      \fmfforce{230,10}{b}
      \end{fmfgraph*}
\end{fmffile}
\caption{The $\rho^{-}$ decay with the $\pi\eta$ mixing in the final state}
\label{fig2}
\end{center}
\end{figure}
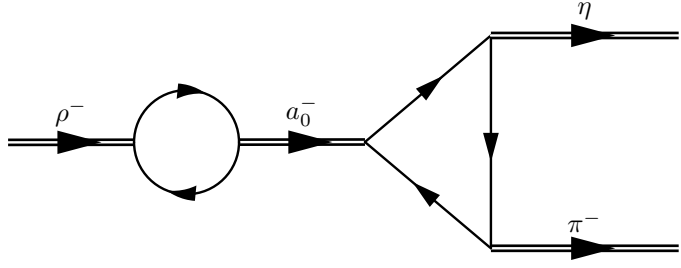
\begin{figure}[h]
\begin{center}
\begin{fmffile}{rho2}
      \begin{fmfgraph*}(250,100)
	      \fmfpen{thin}\fmfleftn{l}{2}\fmfrightn{r}{2}

	      \fmfright{b,a}
	      \fmfleft{f}
	      \fmf{dbl_plain_arrow,lab.side=left,label=$a^{-}_{0}$}{v1,v4}
	      \fmf{dbl_plain_arrow,lab.side=left,label=$\rho^{-}$}{f,v2}
	      \fmf{fermion,left,tension=.6}{v1,v2}
	      \fmf{fermion,left,tension=.6}{v2,v1}

 	      \fmf{fermion,tension=0.5}{p1,v4,p2}
 	      \fmf{fermion}{p2,p1}	      

	      \fmf{dbl_plain_arrow,lab.side=left,label=$\eta$}{p2,a}
 	      \fmf{dbl_plain_arrow,lab.side=left,label=$\pi^{-}$}{p1,b}

	      \fmfforce{180,90}{p2}
	      \fmfforce{180,10}{p1}

	      \fmfforce{250,90}{a}
	      \fmfforce{250,10}{b}
      \end{fmfgraph*}

\end{fmffile}
\caption{The $\rho^{-}$ decay though transition into the $a^{-}_0$ meson}
\label{fig3}
\end{center}
\end{figure}
The first diagram describes the amplitude which contains the $\pi^{0} \to \eta$ transitions in the final state
\begin{equation}
T_1 = g_{\rho}\epsilon_{\pi\eta}(p_{-}^{\mu} - p_{0}^{\mu})\rho^{-}_{\mu}\eta\pi^{-}\,,
\end{equation}
where $g_{\rho} \approx 6.14$ is defined in~\cite{pepan86}.
The second diagram describes the amplitude containing the intermediate $a_{0}^{-}$ meson
\begin{equation}
T_2 = 2Zg_{\rho}\frac{m_u(m_d-m_u)}{m_{a_0}^2-m_{\rho}^2}\epsilon_{\eta}p^{\mu}\rho^{-}_\mu\eta\pi^{-}\,,
\end{equation}
where the vertex $a^{-}_0\to\eta\pi^{-}$ was defined in~\cite{Volkov:1998ax}
\begin{eqnarray}
\frac{4}{\sqrt{6}}Zg_\rho m_u \epsilon_\eta a^{-}_0\eta\pi^{-}\,, \\
Z = \left(1 - 6\frac{m_u^2}{m_{a_1}^2}\right)^{-1}\,,
\label{a0njl}
\end{eqnarray}
and $m_{a_1} = 1230$ MeV is the mass of the $a_1$ meson~\cite{PDG}.

Thus, for branching fractions we get
\begin{eqnarray}
\mathcal{B}_1 &=& \epsilon_{\pi\eta}^2\frac{\lambda^{3/2}(m_{\rho}^2,m_{\eta}^2,m_{\pi}^2)}{\lambda^{3/2}(m_{\rho}^2,m_{\pi}^2,m_{\pi}^2)} = 1.78 \cdot 10^{-5}\,, \\
\mathcal{B}_2 &=& 4Z^2\epsilon_\eta^2\left(\frac{m_u(m_d - m_u)}{m_{a_0}^2 - m_{\rho}^2}\right)^2\frac{\lambda^{3/2}(m_{\rho}^2,m_{\eta}^2,m_{\pi}^2)}{\lambda^{3/2}(m_{\rho}^2,m_{\pi}^2,m_{\pi}^2)} = 0.33 \cdot 10^{-5}\,,
\end{eqnarray}
where $\lambda(s,m_{\eta(\eta')}^2,m_\pi^2) = (s - m_{\eta(\eta')}^2 - m_\pi^2)^2 - 4m_{\eta(\eta')}^2m_\pi^2$.


We note that in these calculations we take into account only the ground state of $a_0$ because the decay with the intermediate $a_0(1450)$ is suppresed by a large mass of the radial-excited meson.

Our estimates coincide with one taken in~\cite{Nussinov:2008gx}. These estimates do not contradict known experimental limits~\cite{PDG,Ferbel:1966zz}.

\section{The decay $\tau^{-}\to\eta(\eta')\pi^{-}\nu$}
The description of the decay $\tau\to\pi\pi\nu$ was obtained in~\cite{Volkov:2012uh} with satisfactory agreement with current experimental data.

\begin{figure}[h]
\begin{center}
\begin{fmffile}{tau1}
      \begin{fmfgraph*}(250,100)
	      \fmfpen{thin}\fmfleftn{l}{2}\fmfrightn{r}{2}
 
	      \fmfright{b,a1}
	      \fmfleft{f,fb}
	      \fmflabel{$\tau^-$}{fb}
	      \fmflabel{$\nu$}{f}
	      \fmf{fermion}{fb,v1,f}
 	      \fmf{fermion}{p1,v2,p2}
	      \fmf{dbl_plain_arrow,lab.side=left,label=$V^{-}$}{v1,v2}
 	      \fmf{fermion}{p2,p1}	      

 	      \fmf{dbl_plain_arrow,lab.side=left,label=$\pi^{-}$}{p1,b}

	      \fmf{dbl_plain_arrow,lab.side=left,label=$\pi^{0}$}{p2,v5}

	      \fmf{fermion,left,tension=.5}{v5,v6}
	      \fmf{fermion,left,tension=.5}{v6,v5}

	      \fmf{dbl_plain_arrow,lab.side=left,label=$\eta(\eta')$}{v6,a1}

	      \fmfforce{175,90}{p2}
	      \fmfforce{175,10}{p1}

	      \fmfforce{270,90}{a1}
	      \fmfforce{270,10}{b}

	      \fmfforce{0,90}{fb}
	      \fmfforce{0,10}{f}

      \end{fmfgraph*}
\end{fmffile}
\caption{The vector contribution to the $\tau$ decay ($V^{-}$ includes a contact term and terms with the intermediate $\rho(770)$ and $\rho(1450)$ mesons)}
\label{fig4}
\end{center}
\end{figure}
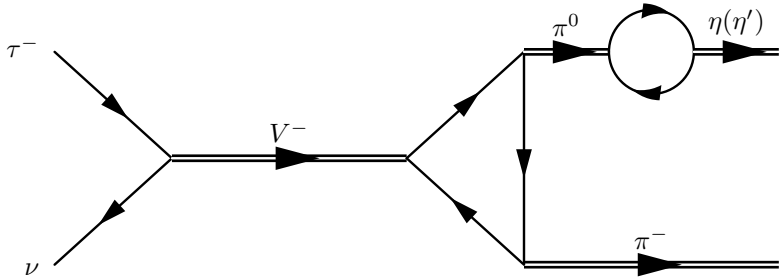

We use the amplitude from~\cite{Volkov:2012uh} with the $\pi^{0} \to \eta(\eta')$ transitions~(\ref{pietatrans}) in the final states (see Fig.~\ref{fig4})\footnote{We neglect the $p^2$ dependence for a rough estimate.}
\begin{equation}
T_V = \epsilon_{\pi\eta(\eta')} m_\rho^2 \left(\left(1-\frac{i\sqrt{q^2}\Gamma_\rho(p^2)}{m_\rho^2}\right)BW_\rho(p^2) + \beta_\rho \frac{p^2}{m_\rho^2} BW_{\rho'}(p^2) \right) (p_{\pi^{-}}^\mu - p_{\eta(\eta')}^\mu) l_\mu \pi^{-} \eta(\eta')\,,
\end{equation}
where the Breit-Wigner relation $BW_{\rho(\rho')}(p^2)$ and $\beta_\rho$ parameter were defined in~\cite{Volkov:2012uh}.
For the processes with the intermediate vector meson we get contributions to branching fractions
\begin{eqnarray}
\mathcal{B}_V(\tau\to\eta\pi\nu) &=& 4.35 \cdot 10^{-6}\,, \\
\mathcal{B}_V(\tau\to\eta'\pi\nu) &=& 1.11 \cdot 10^{-8}\,.
\end{eqnarray}

The $W^{-} \to a_0^{-}$ transition takes the form
\begin{equation}
\frac{\sqrt{3}}{4g_\rho}g_{EW}|V_{ud}|(m_d - m_u)p^{\mu}W_{\mu}^{-}a_0^{-}\,,
\end{equation}
where $g_{EW}$ is the electroweak constant.

\begin{figure}[h]
\begin{center}
\begin{fmffile}{tau2}
      \begin{fmfgraph*}(250,100)
	      \fmfpen{thin}\fmfleftn{l}{2}\fmfrightn{r}{2}

	      \fmfright{b,a}
	      \fmfleft{f,fb}
	      \fmflabel{$\tau^-$}{fb}
	      \fmflabel{$\nu$}{f}
	      \fmf{dbl_plain_arrow,lab.side=left,label=$a_0^{-}$}{v1,v4}
	      \fmf{boson,label=$W^{-}$}{v2,v3}
	      \fmf{fermion,left,tension=.5}{v1,v2}
	      \fmf{fermion,left,tension=.5}{v2,v1}
	      \fmf{fermion}{fb,v3,f}
 	      \fmf{fermion,tension=0.5}{p1,v4,p2}
 	      \fmf{fermion,lab.side=left}{p2,p1}	      

	      \fmf{dbl_plain_arrow,lab.side=left,label=$\eta(\eta')$}{p2,a}
 	      \fmf{dbl_plain_arrow,lab.side=left,label=$\pi^{-}$}{p1,b}

	      \fmfforce{205,90}{p2}
	      \fmfforce{205,10}{p1}

	      \fmfforce{270,90}{a}
	      \fmfforce{270,10}{b}

	      \fmfforce{0,90}{fb}
	      \fmfforce{0,10}{f}

      \end{fmfgraph*}

\end{fmffile}
\caption{The scalar contribution to $\tau$ decay}
\label{fig5}
\end{center}
\end{figure}
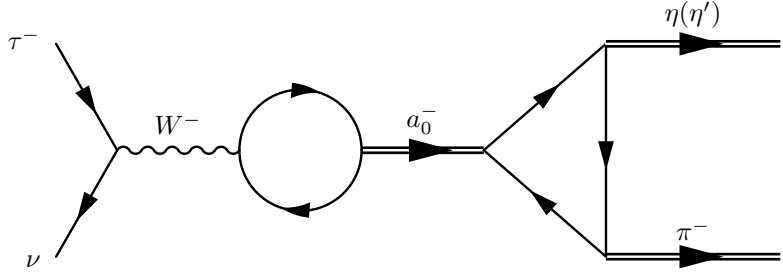

 

The amplitude with the intermediate scalar meson (see Fig.~\ref{fig5}) takes the form
\begin{equation}
T_S = 2Zm_u(m_d-m_u)\epsilon_{\eta(\eta')}(BW_{a_0}(p^2) + \beta_{a_0\eta(\eta')\pi}BW_{a_0'}(p^2))p^{\mu}l_\mu\pi^{-} \eta(\eta')
\label{scalamp}\,,
\end{equation}
where $BW_{a_0(a_0')}(p^2)$ is the Breit–Wigner formula for the $a_0(a_0')$ meson with $m_{a_0} = 980$ MeV, $m_{a_0'} = 1474$ MeV, $\Gamma_{a_0'}(m_{a_0'}) = 265$ MeV taken from PDG~\cite{PDG} and $\Gamma_{a_0}(m_{a_0}) = 100$ MeV calculated from~(\ref{a0njl}) which coincides with the upper PDG limit~\cite{PDG}.
For the estimation of the contribution of the radial-excited $a_0^{-}(1450)$ to the $\tau$ decays we should use the extended NJL model~\cite{VolkovWeiss,yadPh,Volkov:2002iz}. The amplitudes $A_{a_0'\to\eta(\eta')\pi}$ of the $a_0'\to\eta(\eta')\pi$ decays can be found in~\cite{Volkov:2002iz}. The transition $W^{-} \to a^{-}(1450)$ takes the form
\begin{equation}
C_{Wa_0'} = \frac{\sqrt{3}}{4g_\rho}g_{EW}|V_{ud}|(m_d - m_u)\left(\frac{\cos(\phi+\phi_0)}{\sin(2\phi_0)} + \Gamma\frac{\cos(\phi - \phi_0)}{\sin{(2\phi_0)}}\right)p^{\mu}W_{\mu}^{-}a_0^{-}\,,
\end{equation}
where $\phi_0 = 65.5^{\circ}$ and $\phi = 72.0^{\circ}$ are the mixing angles, and $\Gamma = 0.54$.

Thus, we get the $\beta_{a_0\eta(\eta')\pi}$ parameter
\begin{equation}
\beta_{a_0\eta(\eta')\pi} = e^{i\pi}C_{Wa_0'}\frac{\sqrt{6}}{4Z}\frac{A_{a_0'\to\eta(\eta')\pi}}{m_u}\,,
\end{equation}
where phase factor $e^{i\pi}$ are taken similarly~\cite{Volkov:2012uh}. The values $\beta_{a_0\eta\pi} = -0.24$ and $\beta_{a_0\eta'\pi} = -0.26$ do not contradict with ones given in~\cite{Paver:2011md,Nussinov:2009sn}.
The contributions to the branching fractions from the amplitude~(\ref{scalamp}) are
\begin{eqnarray}
\mathcal{B}_S(\tau\to\eta\pi\nu) &=& 0.37 \cdot 10^{-6}\,, \\
\mathcal{B}_S(\tau\to\eta'\pi\nu) &=& 2.63 \cdot 10^{-8}\,.
\end{eqnarray}

The expression for the total width is
\begin{eqnarray}
\Gamma &=& \frac{G_f^2|V_{ud}|^2}{384\pi m_{\tau}^2}\int_{m_{\eta(\eta')}^2 + m_\pi^2}^{m_\tau^2} \frac{\dd s}{s^3}\lambda^{1/2}(s,m_{\eta(\eta')}^2,m_\pi^2)(m_\tau^2 - s)^2 \nn \\
&\times& \left(|T_V|^2(2s+m_\tau^2)\lambda(s,m_{\eta(\eta')}^2,m_\pi^2) + |T_S|^23m_\tau^2(m_{\eta(\eta')}^2-m_\pi^2)^2\right)\,.
\end{eqnarray}
Note that there is no interference between the vector and scalar intermediate state contributions. Thus, for branchings we get
\begin{eqnarray}
\mathcal{B}(\tau^{-}\to\eta\pi^{-}\nu) &=& 4.72 \cdot 10^{-6}\,, \\
\mathcal{B}(\tau^{-}\to\eta'\pi^{-}\nu) &=& 3.74\cdot 10^{-8}\,.
\end{eqnarray}
Let us note that our estimations for scalar contributions are much less than ones in previous works.




\section{Conclusions}
Our calculations are in qualitative agreement with the previous theoretical estimates obtained in~\cite{Nussinov:2008gx,Nussinov:2009sn,Paver:2010mz,Paver:2011md}.
However, the NJL model allows us to describe the transitions $\pi^{0} \to \eta(\eta')$ and $\rho^{-}(W^{-}) \to a_0^{-}$ using the same methods. As the result, we can compare the contribution of amplitudes with intermediate scalar and vector mesons from uniform positions.
These calculations show that in the decays $\rho^{-}\to\eta\pi^{-}$ and $\tau^{-}\to\eta\pi^{-}\nu$ the scalar meson plays a insignificant role. However, in the decay $\tau^{-}\to\eta'\pi^{-}\nu$ the processes with intermediate $a_0$ and $a_0'$ make contributions comparable with the contributions of intermediate vector mesons.

It is worth noticing that the width of the decay $\tau^{-}\to a^{-}_0\nu$ calculated in the NJL model is close to the values obtained in~\cite{Nussinov:2009sn}
\begin{eqnarray}
&& \Gamma = \frac{G_F^2|V_{ud}|^2m_{\tau}^3}{16\pi}\left(\frac{\sqrt{6}}{2}\frac{m_d-m_u}{g_\rho}\right)^2\left(1 - \frac{m_{a_0}^2}{m_\tau^2}\right)^2\,, \\
&& \mathcal{B}(\tau^{-}\to a^{-}_0\nu) = 3.28 \cdot 10^{-6}\,.
\end{eqnarray}
This confirms the relevance of our expression for the vertex $\tau a_0\nu$ used in~(\ref{scalamp}). For the vertex $a_0\to\eta\pi$ the expression was used~(\ref{a0njl}). We get the amplitude~(\ref{scalamp}) by the matching these expressions through propagator of scalar $a_0$ meson. This contradicts the VDM-like ansatz for the intermediate resonance used in~\cite{Bramon:1987zb,Paver:2010mz,Paver:2011md}
\begin{equation}
\frac{\epsilon_{\pi\eta}^2M_R^2}{M_R^2 - p^2 - iM_R\Gamma_R(p^2)}
\end{equation}
On the other hand, if we use this ansatz for a vector to scalar transition taken in~\cite{Bramon:1987zb,Paver:2010mz,Paver:2011md} and calculate $\rho^{-}\to\eta\pi^{-}$ with this ansatz then we get by an order of magnitude
\begin{equation}
\mathcal{B} \sim \epsilon_{\pi\eta}^2\left(\frac{m_{a_0}^2}{m_{a_0}^2 - m_{\rho}^2}\right)^2\frac{\lambda^{3/2}(m_{\rho}^2,m_{\eta}^2,m_{\pi}^2)}{\lambda^{3/2}(m_{\rho}^2,m_{\pi}^2,m_{\pi}^2)} \sim 10^{-3}\,.
\end{equation}
This estimate for the branching fraction is close to the current experimental limit~\cite{PDG,Ferbel:1966zz} and can be tested in the near future at the high-luminosity $e^{+}e^{-}$ colliders in Novosibirsk and Beijing, for example. Therefore, the problem of relevancy of vector -- scalar transition representation can be clarified.

\section*{Acknowledgments}
We are grateful to B.~A.~Arbuzov, A.~B.~Arbuzov and E.~A.~Kuraev for useful discussions. 
This work was supported by the RFBR grant 10-02-01295-a.

\end{document}